\documentstyle[12pt]{article}

\author{Berezovsky S V\\
Scientific and Technological Center of Electrophysics \\
National Academy of Sciences of Ukraine\\
28 Chernyshevsky St., P.O.BOX 8812 \\
UA-310002 Kharkov, Ukraine\\
Tel.: +380-572-404-720; Fax: +380-572-475-261 \\
E-mail: ipct@pem.kharkov.ua}
\title{Soliton regime in the model with no Lifshitz invariant
}
\date{10 August 1999
}
\input tcilatex

\QQQ{Language}{
American English
}

\begin{document}

\maketitle
\begin{abstract}
Nonlinear properties of the order parameter modulation wave in such systems
as thiourea are described in the framework of the phenomenological model
with no Lifshitz invariant. It is also shown that for some values of the
thermodynamic potential parameters the lock-in transition is continuous in
the type II systems. Close to this transition the soliton lattice appears.

\vskip 0.1cm


\vskip 0.2cm

{\bf PACS} number(s): 64.70.Kb; 05.07.Fh; 64.60.My.
\end{abstract}

\section{Introduction}

When describing the incommensurate (IC) phases in the ferroelectrics of so
called type II \cite{Bruce} the phenomenological approach proposed by Y
Ishibashi and H Shiba \cite{Ishibashi1978} is widely used \cite{ToledanoBook}%
.

In according to this model, the system thermodynamic potential can be
written in the form \cite{Singapore}: 
\begin{equation}
\label{eqFunc}\Phi =\Phi _0\cdot \frac 1L\int\limits_0^L[(\varphi ^{\prime
\prime })^2-g(\varphi \varphi ^{\prime })^2-\gamma (\varphi ^{\prime
}{})^2+q\varphi ^2+\frac p2\varphi ^4+\frac h3\varphi ^6]dx 
\end{equation}
where $\varphi (x)$ is a one-dimensional order parameter (e.g., a component $%
P_y(x)$ of the spontaneous polarization ${\bf P\equiv }P_y$); $\varphi
^{\prime }(x)\equiv \partial \varphi /\partial x,$ $L$ is a crystal length
in the direction of spatial modulation of the order parameter.

In the expression (\ref{eqFunc}) the scale transformation is made \cite
{Singapore} in order to emphasize the physically relevant material
parameters $g,$ $q,$ $p$ ($h=0,1;$ $\gamma \equiv 1$ - like in the \cite
{Singapore} we remain the notation '$\gamma $' to indicate the contribution
of invariant $(\varphi ^{\prime })^2$ which favors the appearance of IC
state). For all known ferroelectrics of type II (sodium nitrite $NaNO_2$ 
\cite{Qiu}, thiourea $SC(NH_2)_2$ \cite{Ishibashi1978}, $Sn_2P_2Se_6$ \cite
{VysochanskiiFerro} and betaine calcium chloride dihydrate ($BCCD$) \cite
{Toledano1995}) the parameter $g$ is negative: $g<0$. For sodium nitrite and 
$Sn_2P_2Se_6$ the direct (virtual) disordered-to-commensurate phase
transition is regarded to be of the first order, thereby $p<0$ and $h=1$ for 
$NaNO_2$ \cite{Ishibashi1981} and $Sn_2P_2Se_6$ \cite{VysochanskiiFerro}. In
the case of thiourea and $BCCD$ the parameter $p>0$ and $h=0$. It is usually
assumed that the only parameter $q$ is dependent on temperature $T$: $%
q=q_0\cdot (T-T_0)$, where $q_0$ and $T_0$ are some constants.

The model (\ref{eqFunc}) fairly well describes a lot of properties of the IC
phase in the ferroelectrics of type II \cite{ToledanoBook}. One of such
properties is a predominantly sinusoidal character of the order parameter
modulation wave. Experimental studies indicate that for the type II
ferroelectrics higher order satellites are of low intensity even in the
close vicinity of the lock-in transition (see for review \cite{Cummins}).
This circumstance is a reason why the one-harmonic approximation is often
used when considering the IC order parameter configuration in, e.g., sodium
nitrite or $Sn_2P_2Se_6$ \cite
{Ishibashi1978,Qiu,VysochanskiiFerro,Toledano1995} 
\begin{equation}
\label{eqSin}\varphi (x)=a\cdot \sin (bx). 
\end{equation}

On the other hand, the recent experiments \cite{Thiourea,Thiourea1} reveal
that for thiourea the dependence of order parameter on position $x$ contains
relatively large contribution of the higher harmonics and can not be
regarded as pure sinusoidal.

The solitonic properties of the modulation functions in thiourea are
explained in \cite{Aramburu}. In order to describe nonlinear features of the
IC order parameter configuration the authors of \cite{Aramburu} make use
another than (\ref{eqFunc}) phenomenological approach \cite{Levanyuk1976}.

The approach \cite{Levanyuk1976} is similar to the model developed for the
systems with two-component order parameter for which the Lifshitz invariant
can be introduced \cite{Levanyuk1976A}. Examples of such systems are
compounds of the $A_2BX_4$ family, e.g., $Rb_2ZnCl_4$ \cite{Cummins}.
Although in the case of type II ferroelectrics the order parameter is
usually considered as one-dimensional, the role of second component is
performed by some other normal coordinate $\xi (x)$ (e.g., $xy$-component of
the elastic strain tensor) \cite{Levanyuk1976}. The function $\xi (x)$
transforms like the first derivative of the order parameter: $\xi (x)\sim
\varphi ^{\prime }(x).$ These transformation properties of $\xi (x)$ allow
to construct the thermodynamic potential with a term analogous to Lifshitz
invariant \cite{Levanyuk1976}.

The approach \cite{Levanyuk1976} is expected to be more appropriate than the
model (\ref{eqFunc}) when interpreting nonsinusoidal configurations of the
order parameter \cite{Aramburu}. To some extent, these expectations are
grounded on the analogy between the approach \cite{Levanyuk1976} and the
theory developed in \cite{Levanyuk1976A}. Really, the latter constitutes a
powerful tool for description of the soliton structures in the compounds of $%
A_2BX_4$ family \cite{Sannikov1980}.

In the present paper we show that the nonlinear contribution to the
modulation wave observed in the experiments for thiourea can be explained in
the framework of the model (\ref{eqFunc}) as well.

Our consideration is based on the nonlinear approximation for the IC order
parameter configuration proposed in \cite{Singapore}. Using it we obtain
that if the system material parameters $g$ and $p$ are so that $g\cdot
p^{-1}\approx -6.0$ ($h=0$, $p>0$) then the ratio $a_3/a_1$ for the
amplitudes of third ($a_3$) and fundamental ($a_1$) harmonics of the
modulation wave is equal to $a_3/a_1\approx 0.1$. This result completely
agrees with the experimental data for thiourea \cite{Thiourea,Thiourea1}.

Moreover, it is found that for some values of the parameters $g$ and $p$
(for systems with $h=0$, $p>0$ these values satisfy the relationship $\gamma
\cdot gp^{-1}\approx -16$) the transition from the IC phase into the
commensurate state is continuous. The order parameter configuration is
domain-like in the proximity of such transition.

The structure of present paper is following. In section 2 we formulate main
features of the order parameter approximation proposed in \cite{Singapore}.
In section 3 we consider the nonlinear configurations of order parameter in
the case of thiourea. The estimations of nonlinear properties for other
known ferroelectrics of type II are given as well. The principal possibility
of existence of the strong soliton regime in the type II systems is
investigated in section 4. The comparison of the obtained results with the
properties of type I systems is presented in section 5.

\section{Sn-approximation for the IC order parameter configuration}

In according with the analysis made in \cite{Singapore}, the equilibrium
configuration of order parameter in the IC phase for the type II systems (%
\ref{eqFunc}) can be approximated as 
\begin{equation}
\label{eqSn}\varphi (x)=a\cdot \limfunc{sn}(bx,k) 
\end{equation}
where $\limfunc{sn}(x,k)$ is the Jacobi elliptic sinus \cite{Abramowitz}. In
(\ref{eqSn}) the amplitude $a$, the wave number $b$ and the elliptic modulus 
$k$ ($0\leq k\leq 1$) are defined by the minimization of the thermodynamic
potential (\ref{eqFunc}) in respect to $a,$ $b,$ $k$ \cite{Singapore}.

In contrast to the approach (\ref{eqSin}), the approximation (\ref{eqSn})
allows to consider not only the linear regime of the IC phase but the
nonlinear configurations of order parameter as well.

Really, if the elliptic modulus $k$ is small ($k\approx 0$) the function (%
\ref{eqSn}) is closed to the dependence (\ref{eqSin}): $\limfunc{sn}%
(x,k\approx 0)\approx \sin (x)$ \cite{Abramowitz}. But when $k\rightarrow 1$
the spatial behavior of elliptic sinus becomes domain-like: the wide regions
with almost constant values $\varphi (x)\approx \pm \varphi _0$ are
separated by the narrow regions where the function (\ref{eqSn}) changes
abruptly.

In the model (\ref{eqFunc}), (\ref{eqSn}) the elliptic modulus $k$ is equal
to zero at the point $q_I=\frac 14\gamma ^2$ of the
disordered-to-incommensurate phase transition \cite{Singapore}. With
decreasing temperature the elliptic modulus $k$ grows and takes its largest
value $k_c$ at the lock-in transition point $q_c$. The preliminary
investigations show that $k_c$ can be close to unity for some values of the
material parameters. For example, if $g=-10,$ $p=1,$ $h=0$ then $k_c=0.965$ 
\cite{Singapore}.

Another important property of the order parameter approximation (\ref{eqSn})
is an additional (in comparison with the approach (\ref{eqSin})) mechanism
causing the change of modulation period.

The continuous dependence of modulation period $P$ on temperature is one of
the most characteristical features of the IC phases \cite{ToledanoBook}. In
the framework of approach (\ref{eqSin}) the period of IC structure is equal
to $P=2\pi /b$. The wave number $b$ depends on temperature only when the $%
(\varphi \varphi ^{\prime })^2$ - invariant is present in the expansion of
the thermodynamic potential (\ref{eqFunc}) \cite{Ishibashi1978,Qiu}: 
\begin{equation}
\label{eqSinB2}b^2=\frac 12\gamma +\frac 18g\cdot a^2. 
\end{equation}
And if the material parameter $g$ is negative then the period $P$ increases
with decreasing temperature as it is observed in experiments \cite{Cummins}.

The period of IC order parameter configuration (\ref{eqSn}) is defined as $%
P=4K(k)/b$ ($K(k)$ is the complete elliptic integral of the first kind \cite
{Abramowitz}). It depends not only on the value of $b$ (as it takes place in
the one-harmonic approach (\ref{eqSin})) but also on the elliptic modulus $k$
which characterizes the degree of nonlinearity of the modulation wave. As
consequence, the approximation (\ref{eqSn}) imposes less strong requirements
on the material parameters, in particular, on $g$ (e.g., the period $P$
grows even when $g=0$ \cite{Singapore,Studies}). Moreover, if the elliptic
modulus $k$ is close to unity $k\rightarrow 1$ (i.e. $K(k)\rightarrow \infty 
$ \cite{Abramowitz}) then the nonlinear mechanism of the increase of
modulation period becomes dominant and $P$ can be very large: $P\rightarrow
\infty $.

The numerical investigation of variational equation for the functional (\ref
{eqFunc}) \cite{StudiesSchet} shows that the approximation (\ref{eqSn})
correctly reproduces nonlinear properties of the equilibrium configuration
of order parameter in the IC phase.

\section{Nonlinear configurations of the order parameter in thiourea}

Now let us to apply the model (\ref{eqFunc}), (\ref{eqSn}) to thiourea.

As it has been mentioned above, in the case of thiourea the material
parameter $h$ equals to zero. For the sake of simplicity we also assume that 
$p=1$.

Numerical minimization of the thermodynamic potential (\ref{eqFunc}), (\ref
{eqSn}) in respect to $a,$ $b,$ $k$ shows that if the material parameter $g$
is equal to $g=-6.0$ then the elliptic modulus takes the value $k_c\approx
0.923$ at the point $q_c=-0.290$ of the lock-in transition. Using the
formulae for the Fourier expansion of elliptic sinus \cite{Abramowitz} or
the procedures of fast Fourier transformation (FFT), one can find that for
this $k_c$ the ratio of amplitudes of the third ($a_3$) and the fundamental (%
$a_1$) harmonics of the modulation wave (\ref{eqSn}) is equal to $%
a_3/a_1\approx 0.104$ and also $a_5/a_1\approx 0.012$ ($a_5$ is an amplitude
of the fifth harmonic). The spatial dependence of the order parameter $%
\varphi (x)$ at the temperature $q_c$ is shown in figure 1 (full line). The
obtained results are in a good agreement with the experimental data \cite
{Thiourea,Thiourea1}.

Besides the order parameter $\varphi (x)$, some other modulation functions
are also discussed in the case of thiourea \cite{Aramburu}. If such function 
$\xi (x)$ is coupled with the order parameter $\varphi (x)=a\limfunc{sn}%
(bx,k)\equiv a\sin [\theta (x)]$ by the relation $\xi (x)\sim \cos [\theta
(x)]\equiv \limfunc{cn}(bx,k)$ then the dependence of $\xi (x)$ on position $%
x$ has the form depicted in figure 1 by broken line (cf. with figure 3 in 
\cite{Aramburu}).

Of course, the complete description of $\xi (x)$ and other modulation
functions requires a modification of the thermodynamic potential (\ref
{eqFunc}) by including additional invariants which correspond to
interactions of $\xi (x),...$ with each other and with the order parameter.
Such an analysis is, however, beyond the scope of the present paper.

The nonlinear properties of the IC modulation wave in thiourea essentially
differ from ones for sodium nitrite. Taking for estimations the material
parameters given in \cite{Qiu} (in our notations they correspond to $%
g=-9.51, $ $p=-0.651,$ $h=1$) we find that for $NaNO_2$ the lock-in value of
elliptic modulus is $k_c\approx 0.589,$ $a_3/a_1\approx 0.026$ and $%
a_5/a_1\approx 0.0007$. Note that these results are very close to ones
obtained analytically in \cite{Levanyuk1990}.

As for other ferroelectrics of type II, estimations reveal that for $%
Sn_2P_2Se_6$ ($g=-1.37,$ $p=-0.19,$ $h=1$ \cite{VysochanskiiFerro}) the
characteristics of modulation wave at the lock-in transition point are the
following: $k_c\approx 0.706,$ $a_3/a_1\approx 0.041,$ $a_5/a_1\approx
0.0018;$ and for $BCCD$ ($g=-8.0,$ $p=2.0,$ $h=0$ \cite{Toledano1995}) - $%
k_c\approx 0.887,$ $a_3/a_1\approx 0.086$ and $a_5/a_1\approx 0.0082$.

\section{Soliton regime in the type II systems}

Results given in the preceding section describe the nonlinear properties of
the IC modulation in four known compounds belonging to the type II
ferroelectrics. However, these estimations do not answer on more general
question: to what extent the soliton regime can develop in the model (\ref
{eqFunc}), (\ref{eqSn}) in principle. In order to clarify the situation we
have investigated the behavior of systems (\ref{eqFunc}), (\ref{eqSn}) in
the limit $k\rightarrow 1$ in more detail.

When $k\approx 1$ the thermodynamic potential (\ref{eqFunc}), (\ref{eqSn})
acquires the form \cite{Singapore}: 
\begin{equation}
\label{FuncK1} 
\begin{array}[t]{ll}
\Phi \approx & a^2\left[ q\left( 1+\frac 12k^{\prime 2}\right) -\Lambda
^{-1}\left( q+\frac 23\gamma b^2-\frac 8{15}b^4\right) -\frac 12k^{\prime
2}\Lambda ^{-1}\left( q+\frac 8{15}b^4\right) \right] + \\  
& a^4\left[ \frac 12p\left( 1+k^{\prime 2}\right) -\Lambda ^{-1}\left( \frac
23p+\frac 2{15}gb^2\right) -k^{\prime 2}\Lambda ^{-1}\left( \frac 23p+\frac
1{15}gb^2\right) \right] + \\  
& a^6\cdot \frac 13h\cdot \left( 1+\frac 32k^{\prime 2}-\frac{23}{15}\Lambda
^{-1}-\frac{23}{10}k^{\prime 2}\Lambda ^{-1}\right) 
\end{array}
\end{equation}
where $k^{\prime 2}=1-k^2,$ $\Lambda $$=\ln (4/k^{\prime });$ $\Lambda
^{-1}\rightarrow 0$ if $k\rightarrow 1$.

The equilibrium wave number $b$ can be found from the equation $\partial
\Phi /\partial b=0,$ what yields (cf. with (\ref{eqSinB2})): 
\begin{equation}
\label{eqSnB2}b^2=\frac 58\gamma \cdot (1+\frac 12k^{\prime 2})+\frac
18g\cdot a^2\cdot (1+k^{\prime 2}). 
\end{equation}

Taking into account (\ref{eqSnB2}) the thermodynamic potential (\ref{FuncK1}%
) can be rewritten as 
\begin{equation}
\label{FunK1A} 
\begin{array}[t]{ll}
\Phi \approx & a^2\cdot \left( 1+\frac 12k^{\prime 2}\right) \cdot \left[
q-\Lambda ^{-1}\left( q+\frac 5{24}\gamma ^2\right) \right] + \\  
& a^4\cdot \left( 1+k^{\prime 2}\right) \cdot \left[ \frac 12p-\Lambda
^{-1}\left( \frac 23p+\frac 1{12}\gamma g\right) \right] + \\  
& a^6\cdot \left( 1+\frac 32k^{\prime 2}\right) \cdot \left[ \frac
13h-\Lambda ^{-1}\left( \frac{23}{45}h+\frac 1{120}g^2\right) \right] 
\end{array}
\end{equation}

The function $\Lambda ^{-1}(k)$ changes much slower then $k^{\prime 2}$
approaches to zero (e.g., for $k^{\prime 2}=0.1$ the value of $\Lambda
^{-1}(k)$ is $0.394$, for $k^{\prime 2}=4\times 10^{-8}$ - $\Lambda
^{-1}(k)\approx 0.1$). Hence, it is reasonable to omit the terms
proportional to $k^{\prime 2}$ in (\ref{FunK1A}).

The further analysis depends on the value of material parameter $h.$ For the
systems with $h=0$, $p>0$ the results are as follows.

The equation $\partial \Phi /\partial a=0$ defines the equilibrium amplitude 
$a$. Substituting its solution into the expression (\ref{FunK1A}) and
comparing the result with the thermodynamic potential for the commensurate
state $\Phi _c=-q^2/2p$, we find the effective temperature $q_c$ of the
lock-in transition: 
\begin{equation}
\label{eqQc} 
\begin{array}{c}
q_c=5g^{-2}p^2\cdot \left[ \gamma \cdot gp^{-1}-4+4\left( 1-\frac 12\gamma
\cdot gp^{-1}\right) ^{1/2}\right] \approx \\ 
\approx \frac 52\gamma ^2\cdot \left( \gamma \cdot gp^{-1}-4\right) . 
\end{array}
\end{equation}
In (\ref{eqQc}) the approximate (right hand side) formula is derived for the
case when the term proportional to $a^6$ is neglected in (\ref{FunK1A}).
This expression clearly demonstrates main features of the dependence of $q_c$
on $gp^{-1}$.

Now we consider the conditions under which the elliptic modulus can be equal
to unity at the lock-in transition point $q_c$.

The thermodynamic potential (\ref{FunK1A}) depends on the elliptic modulus $%
k $ only through the function $\Lambda ^{-1}(k)$ (remember we omit the terms
proportional to $k^{\prime 2}$ in (\ref{FunK1A})). Thereby it is convenient
to formulate the variational task for $k$ in terms of $\Lambda ^{-1}.$ In
these terms the condition $k_c=1$ means $\Lambda ^{-1}(k_c)=0$, and the
equation $\partial \Phi /\partial k=0$ is equivalent to $\partial \Phi
/\partial (\Lambda ^{-1})=0$. Solving the latter at the temperature $q_c$ (%
\ref{eqQc}), we find that the function $\Lambda ^{-1}(k_c)$ is equal to zero
if the material parameters satisfy the relation: 
\begin{equation}
\label{eqGP}\gamma \cdot gp^{-1}=-16. 
\end{equation}

Therefore, for the systems with material parameters $g$ and $p$ related with
each other in according to (\ref{eqGP}) ($h=0$), the elliptic modulus of the
order parameter modulation wave (\ref{eqSn}) is equal to unity at the point
of lock-in transition. It means that for such systems the transition from
the IC phase into the commensurate state is continuous. Close to this
transition the order parameter spatial configuration is domain-like, and the
soliton density $n_S=\pi /2K(k)$ (see \cite{Aramburu} and references
therein) approaches to zero with decreasing temperature $q$ to its lock-in
value $q_c$.

Numerical calculations confirm the results of analytical investigation (see
table 1).

For the case $gp^{-1}=-16$ the elliptic modulus $k_c$ has approached to
unity, but we have stopped calculations at the value $k_c=0.9999.$ The
properties of elliptic functions change abruptly at the point $k=1$ \cite
{Abramowitz}, and there are some difficulties to reproduce the point $k=1$
numerically.

As it follows from table 1 the dependence of $k_c$ on $gp^{-1}$ has a
maximum when $gp^{-1}=-16$. For little ($gp^{-1}\sim 1$) and large ($%
gp^{-1}\sim 100$) values of the parameter combination $gp^{-1}$ the
contribution of higher harmonics in the modulation wave is relatively small.

The analysis for the case of systems with $h=1$, $p<0$ can be made in the
similar manner. Here we point out only the following.

When $h=1$, $p<0,$ the direct disordered-to-commensurate phase transition is
of the first order. As consequence, for large enough values of the parameter 
$\left| p\right| $ the range of IC phase stability can be relatively small
like it takes place for $NaNO_2$ ($q_c\approx +0.07$) \cite{Qiu}.
Nevertheless, for any $p$ there exists some $g$ for which $k_c=1.$ For
example, for $p=-0.65$ (the case of sodium nitrite \cite{Singapore,Qiu}) the
lock-in value $k_c$ of the elliptic modulus of the modulation wave (\ref
{eqSn}) equals to $1$ if the material parameter $g$ is $g\approx -4.5$.

However, due to the presence of term proportional to $a^6$ in (\ref{eqFunc})
, i.e. due to $h=1$, the material parameters $g$ and $p$ are not so
correlated as in the case of $h=0$ (in fact, when $h=0$ the ratio $g/p$ is
relevant rather than the parameters $g$ and $p$ themselves). As consequence,
for the systems with $h=1,$ $p<0$ the dependence $g(p)$ providing $k_c=1$ is
more complex than (\ref{eqGP}).

\section{Discussion}

In the present paper we have shown that in the framework of the
phenomenological model with no Lifshitz invariant \cite
{Ishibashi1978,Ishibashi1981} different nonlinear configurations of the IC
order parameter can be described: almost sinusoidal one as in the case of
sodium nitrite ($a_3/a_1\leq 0.03$); one with more large contribution of
higher harmonics as in thiourea ($a_3/a_1\approx 0.1$); strong soliton
regimes when the lock-in transition is continuous (such compound is unknown
at the moment). Nonlinear properties of the concrete system are defined by
values of the material parameters $g$ and $p$ (see, e.g., table 1).

The specific role of $(\varphi \varphi ^{\prime })^2$ - invariant should be
emphasized when discussing the nonlinear features of the IC\ order parameter
configurations. If this term is not included in the thermodynamic potential (%
\ref{eqFunc}) the soliton structure does not develop (see also \cite
{Singapore,Studies}). From this point of view, $(\varphi \varphi ^{\prime
})^2$ - term is analogous to the Umklapp invariant of relatively low order
(the anisotropy invariant) which favors the appearance of domain-like
structures in the type I ferroelectrics \cite{ToledanoBook,Levanyuk1976A}.
The difference between $(\varphi \varphi ^{\prime })^2$ - term in (\ref
{eqFunc}) and the anisotropy invariant is that the former is a part of the
gradient gain of the thermodynamic potential and can not influence on
characteristics of the commensurate phase. On the contrary, in the type I
ferroelectrics the behavior of IC and commensurate phases is correlated due
to the anisotropy invariant (it belongs to the local interactions). In the
case of type II systems analogous correlation appears only for some specific
values of the material parameters $g$ and $p$, i.e. only for some correlated
actions of the invariants $(\varphi \varphi ^{\prime })^2$ and $\varphi ^4$
which define the low-temperature behavior of IC and commensurate phases.

The existence of IC state in the type I systems is caused by symmetric
properties (the Lifshitz condition is not fulfilled) \cite{ToledanoBook}.
For the systems of type II such global reasons are absent and the spatial
modulation of the order parameter is a consequence of specific features of
the interatomic interactions \cite{ToledanoBook}.

As result, in the type II ferroelectrics the appearance of soliton regime
has no systematic character, in contrast to the situation which takes place
for, e.g., compounds of the $A_2BX_4$ family.

\vskip 0.8cm

{\bf Acknowledgments}\vskip .3cm

The author would like to thank V F Klepikov and Yu M Vysochanskii for the
fruitful discussion and support.

\begin{center}
\newpage{\bf Figure Captions.}
\end{center}

{\bf Figure 1}. The order parameter $\varphi (x)$ (full line) and the
modulation function $\xi (x)$ (broken line) as functions of position $x$ at
the temperature $q_c=-0.29$ (the lock-in transition point). The material
parameters are the following: $g=-6.0$, $p=1.0$, $h=0$. The amplitudes of
order parameter $\varphi (x)$ and function $\xi (x)$ are arbitrary.

\begin{center}
{\bf Table Caption.}
\end{center}

{\bf Table 1}. Characteristics of the order parameter modulation wave in the
type II systems (the case $h=0$, $p>0$) at the point $q_c$ of lock-in
transition for different values of the material parameters $g$ and $p$: $k_c$
is a lock-in value of the elliptic modulus, $a_3/a_1$ is a ratio of the
third and first harmonics, $n_S$ is a soliton density.

\begin{center}
\begin{tabular}{ccccccc} \hline\cline{1-5}
$g/p$&$q_c$&$k_c$&$a_3 / a_1$&$n_S$\\ \hline
-1.0&-0.690&0.768&0.052&0.81\\ 
-10.&-0.212&0.965&0.136&0.57\\ 
-16.&-0.156&0.9999&0.262&0.27\\ 
-20.&-0.134&0.970&0.142&0.56\\ 
-100.&-0.048&0.766&0.052&0.81\\ 
\hline \cline {1-5}
\end{tabular}
\end{center}

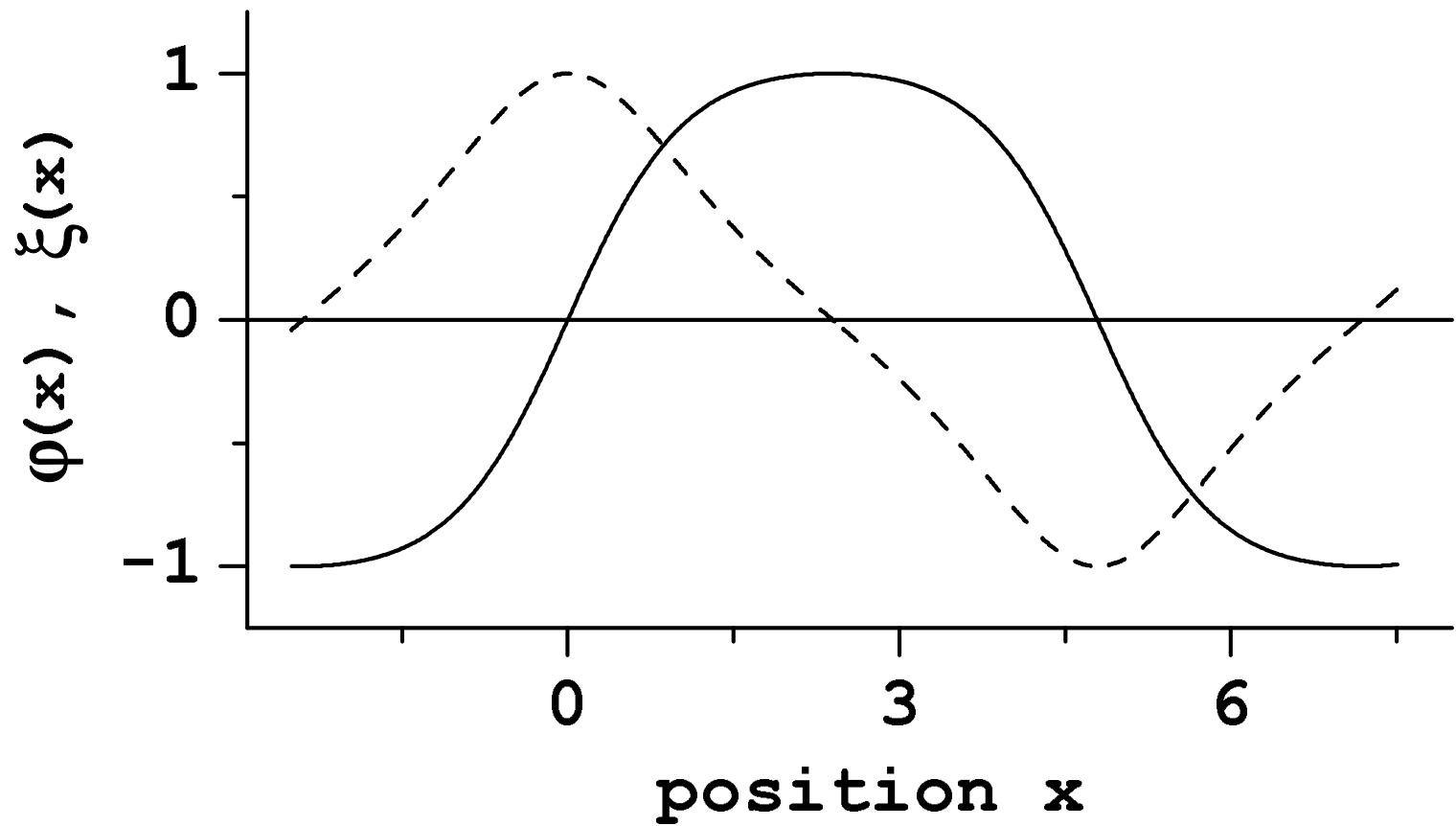
\end{document}